\documentclass[unnumsec,webpdf,traditional,medium]{my-oup-authoring-template}

\onecolumn 

\theoremstyle{thmstyleone}%
\theoremstyle{thmstyletwo}%
\theoremstyle{thmstylethree}%

\begin{document}

\journaltitle{arXiv}
\DOI{10.48550/arXiv.2502.00214}
\copyrightyear{2025}
\pubyear{2025}
\appnotes{Preprint}

\firstpage{1}


\title[Proportional Effect Models]{A critical evaluation of longitudinal proportional effect models}

\author[1,$\ast$]{Michael C. Donohue\ORCID{0000-0001-6026-2238}}
\author[2]{Philip S. Insel\ORCID{0000-0002-6270-5490}}
\author[1]{Oliver Langford\ORCID{0009-0006-8951-1326}}

\authormark{Donohue et al.}

\address[1]{\orgdiv{Alzheimer's Therapeutic Research Institute}, \orgname{University of Southern California}, \orgaddress{\street{9860 Mesa Rim Rd, San Diego}, \state{CA}, \postcode{92121}, \country{USA}}}
\address[2]{\orgdiv{Department of Psychiatry}, \orgname{University of California, San Francisco}, \state{CA}, \country{USA}}

\corresp[$\ast$]{Corresponding author. \href{email:mdonohue@usc.edu}{mdonohue@usc.edu}}




\abstract{Nonlinear longitudinal proportional effect models have been proposed to improve power and provide direct estimates of the proportional treatment effect in randomized clinical trials. These models assume a fixed proportional treatment effect over time, which can lead to bias and Type I error inflation when the assumption is violated. Even when the proportional effect assumption holds, these models are biased, and their inference is sensitive to the labeling of treatment groups. Typically, this bias favors the active group, inflates Type I error, and can result in one-sided testing. Conversely, the bias can make it more difficult to detect treatment harm, creating a safety concern.}



\maketitle

\section{Introduction}

\cite{wang2024novel} provide a tutorial on fitting longitudinal proportional effect models for clinical trials. These models have garnered significant attention in \emph{Statistics in Medicine} due to their promise of direct estimates of proportional treatment effects and increased statistical power \citep{wang2024novel, wang2018novel, raket2022progression}. In this letter, we will illustrate that these models are biased even when model assumptions are correct and explain why we recommend avoiding them.

While most literature on these models has focused on Alzheimer's disease, they have been applied to other therapeutic areas as well. For instance, a joint model of longitudinal and time-to-event data has been proposed for an Amyotrophic Lateral Sclerosis platform trial, as discussed in the tutorial \citep{quintana2023design, wang2024novel}. In this case, the proportional hazard sub-model and the longitudinal sub-model share the same proportional effect.

These longitudinal models seem like novel and powerful adaptations of familiar models like the Cox proportional hazards model. However, they behave differently in several important ways that are not well described in the tutorial or elsewhere in the literature. They are biased, and their inference is sensitive to the labeling of treatment groups. The group labeling suggested in the tutorial can induce bias in favor of active treatment, inflate Type I error, and result in one-sided testing, even when the proportional treatment effect assumption is true. This bias in favor of active treatment also makes it more difficult to detect treatment harm, creating a safety concern.

The models assume that the mean trend over time in the active intervention group is a fixed proportion of the control group mean. This fixed proportional effect parameterization can be written

\begin{equation}
f_A(t)=f_C(t)(1-\theta),\label{eq1}
\end{equation}
where $f_A(t)$ and $f_C(t)$ are the mean response for active and control at time $t$. The control group mean might be estimated using regression splines, for example. The parameterization allows an interpretation of $\theta$ as the proportional effect

\begin{equation}
\theta = \frac{f_C(t)-f_A(t)}{f_C(t)}\nonumber
\end{equation}
for any time $t$, which is thought to facilitate clinical interpretation. Expressing $\theta$ this way shows that it is not invariant to location shifts of $f_C(t)$ and it has no finite definition when $f_C(t)=0$. 

For a cognitive outcome in a progressive stage of disease like mild to moderate dementia, it might be reasonable to assume change from baseline is declining monotonically and avoiding zero. However, a cognitive outcome in the presymptomatic stage of disease is likely to improve before declining  \citep{sperling2023trial}. Even trials in mild cognitive impairment have demonstrated mean improvement before decline, and symptomatic treatments in this population have demonstrated an early treatment benefit that wanes (violating the proportional effect assumption)  \citep{donohue2023natural}.

Longitudinal models with a fixed proportional effect assumption are unreliable in real clinical trial data and simulations when the assumption fails. They have even failed to converge for high-profile trials  \citep{salloway2021trial}, and model estimates are inconsistent with less constrained models with real clinical trial data \citep{donohue2023natural}. Simulation studies have shown Type I error as high as 93\% when the fixed proportional effect assumption is violated, the control group trend is near zero, and an unstructured covariance structure is attempted \citep{donohue2023natural}.

The mixed model repeated measures (MMRM) is a linear model that includes parameters for the control group mean and an additive treatment effect at each visit. We refer to this as an unstructured mean assumption for both control and active groups. The proportional MMRM (pMMRM) is a nonlinear model that assumes an unstructured mean for control and a fixed proportional treatment group difference as defined in (\ref{eq1}). 

The pMMRM is purported to be more powerful than MMRM. If this result extends to the special case of a single follow-up visit, it suggests a proportional test should be more efficient than a two-sample $t$-test. If true, this would be an extraordinary finding.

To better understand the proportional treatment effect parameterization and its efficiency relative to the usual group mean difference, the next section provides simple simulations examining the cross-sectional setting. We find that nonlinear least squares estimation yields treatment effect estimates that are sensitive to the group labeling, and biased in favor of the active group under the labeling suggested in the literature. Conversely, this bias will make it more difficult to detect treatment harm. Furthermore, the bias occurs even when the proportional effect assumption is true. These problems persist in a simulation study with longitudinal data, suggesting the purported efficiency gains are erroneously inflated by bias. Code to reproduce the simulations in R is available from \url{https://github.com/mcdonohue/PropEffects}

\section{Cross-sectional proportional effect model}\label{cross-sectional}

We consider a special case of the pMMRM for a single time point. The cross-sectional proportional treatment effect estimator $\hat\theta$ is defined

\begin{equation}
E(Y\mid x) = \hat\beta_C(1-\hat\theta x),\nonumber
\end{equation}
where $x=0$ for the control group and $x=1$ for the active group. The mean in the control group is estimated by $\hat\beta_C$, and the active group estimate is $\hat\beta_A=\hat\beta_C(1-\hat\theta)$, so that the estimated proportional reduction with active treatment is $\hat\theta = \frac{\hat\beta_C-\hat\beta_A}{\hat\beta_C}$. Following similar notation, the familiar two-sample $t$-test estimator, $\hat\delta$, is defined

\begin{equation}
E(Y\mid x) = \hat\beta_C + \hat\delta x.\nonumber
\end{equation}
The additive treatment effect estimator is simply the difference $\hat\delta =\hat\beta_A - \hat\beta_C$, and in contrast $\beta_C$ is not required to be bounded away from zero.

\begin{figure}
\centering
\includegraphics[scale=0.9]{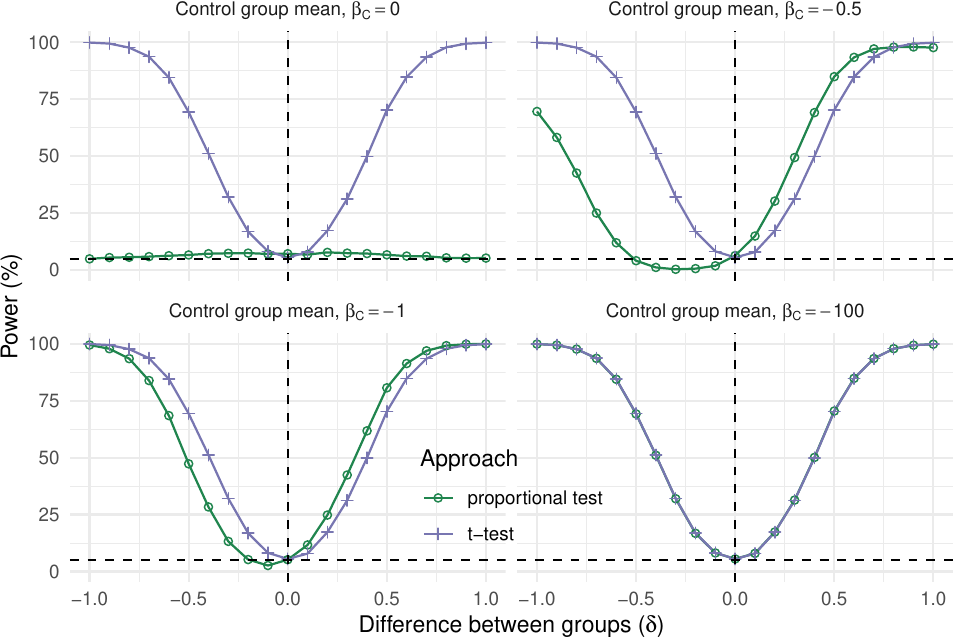}
\caption{\label{fig1} Simulated power (and type I error when $\delta$= 0) for the proportional model and two-sample $t$-test based on 10,000 simulations. The control group mean is $\beta_C$ and $\delta$ is the mean group difference. With $\beta_C = 0$, the proportional effect has no finite definition, and simulated power with the proportional test is about 5\% regardless of $\delta$. With $\beta_C$ near zero ($\beta_C=-0.5$ or $-1$), the proportional test appears to have better power with $\delta>0$, and worse power with $\delta<0$. With $\beta_C$ far from zero ($\beta_C=-100$), power for the two methods is identical and the lines are overlapping. Residual variance was simulated to be one, sample size was 50 per group, and the proportional model was fit with R's nonlinear least squares (\texttt{nls}) function  \citep{bates1992}.}
\end{figure}

A simple simulation comparing statistical power for these cross-sectional models shows that the proportional model is sensitive to the location of the control group, the direction of the effect, and the group labels (Figure \ref{fig1}). The proportional effect model \emph{appears} to improve power, but only with $\delta>0$ and $\beta_C$ near zero. With $\delta<0$, the simple $t$-test generally dominates the proportional test, which means that the proportional test is demonstrating diminished power to detect treatment harm, raising a safety concern. With $\beta_C=-100$ (bottom right) the power is identical with the two tests. With $\beta_C=0$ (top left), the treatment effect is not identifiable with the proportional model, and its power hovers around 5\%.

\begin{figure}
\centering
\includegraphics[width=6in]{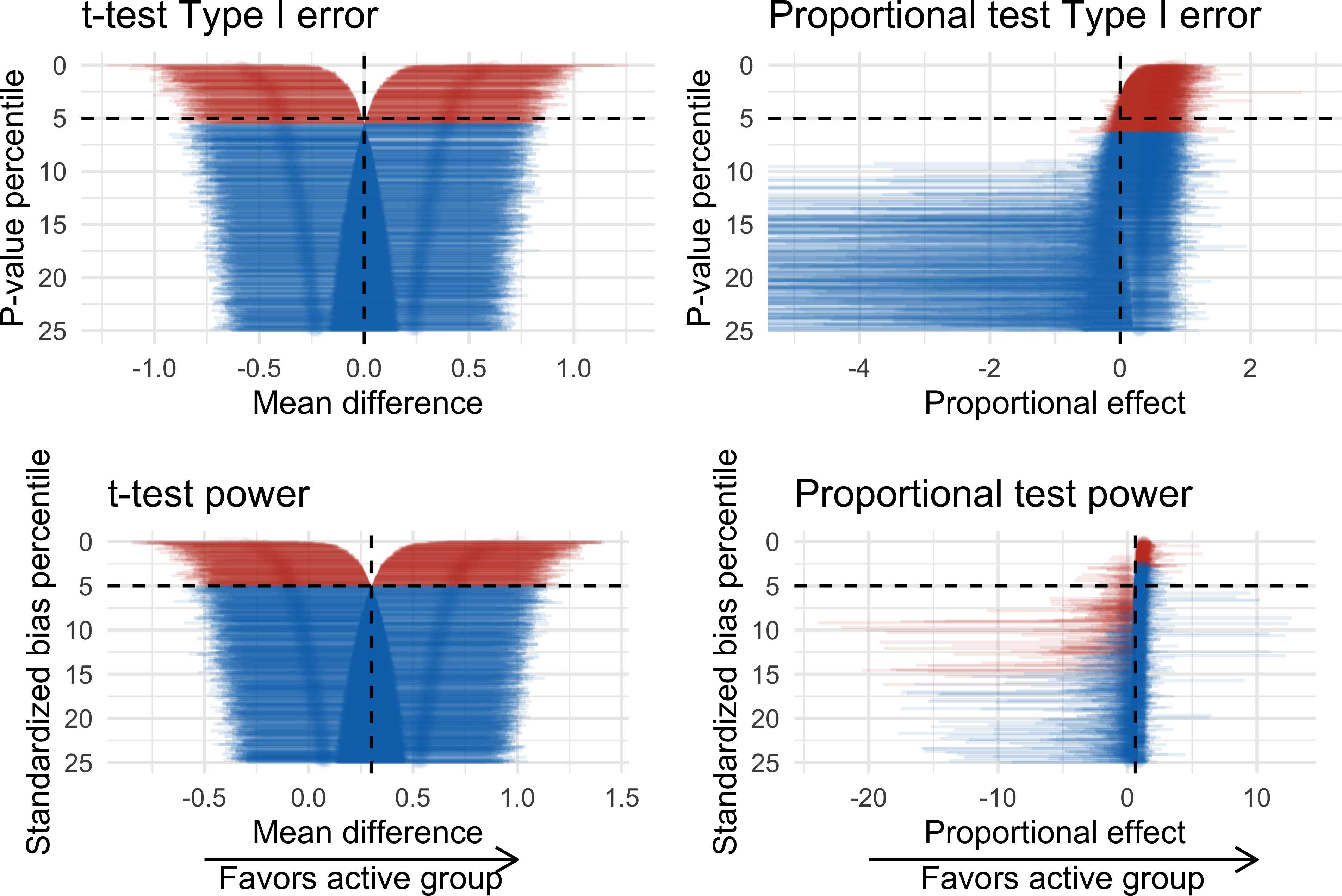}
\caption{\label{figxsectzip} Zipper plots showing estimates and 95\% confidence intervals for cross-sectional simulations with control group mean $\beta_C=-0.5$. The top row has a treatment effect of $\delta=0$ and demonstrates Type I error. The bottom row has a treatment effect of $\delta=0.3$, or equivalently, a proportional effect $\theta = 0.6$. The intervals are sorted so that those associated with the largest standardized bias are toward the top of each panel, and only 25\% of simulations with the largest bias are shown. Red intervals in the top row denote $p<0.05$. Red intervals in the bottom row do not cover the true simulated value (vertical dashed lines). While $t$-test estimates are symmetric about the true effect (left), the plots for the proportional test (right) reveal bias and asymmetric confidence interval widths. Horizontal dashed lines are at the target rejection rate of 5\%. Note that the proportional model p-values are often inconsistent with confidence interval coverage as can be seen with red intervals ($p<0.05$) which cover zero in the upper right panel. Residual variance was simulated to be one, sample size was 50 per group, and the proportional model was fit with R's nonlinear least squares (\texttt{nls}) function  \citep{bates1992}.}
\end{figure}

Further inspection of the simulations shows the apparent improved power with the proportional test when $\beta_C$ is near zero is a result of bias. Figure \ref{figxsectzip} shows that with $\beta_C=-0.5$ and $\delta=0$, the proportional test tends to favor the active group (top right). Bias is also apparent with the proportional test and a treatment effect of $\delta=0.3$, or equivalently, a proportional effect $\theta=0.6$ (bottom right). In some settings, the model behaves like a one-sided test which only rejects the null hypothesis in favor of active treatment. We now explore the implications of this bias for longitudinal models.

\clearpage

\section{Longitudinal proportional effect model}\label{longitudinal}

\begin{figure}
\centering
\includegraphics[scale=0.9]{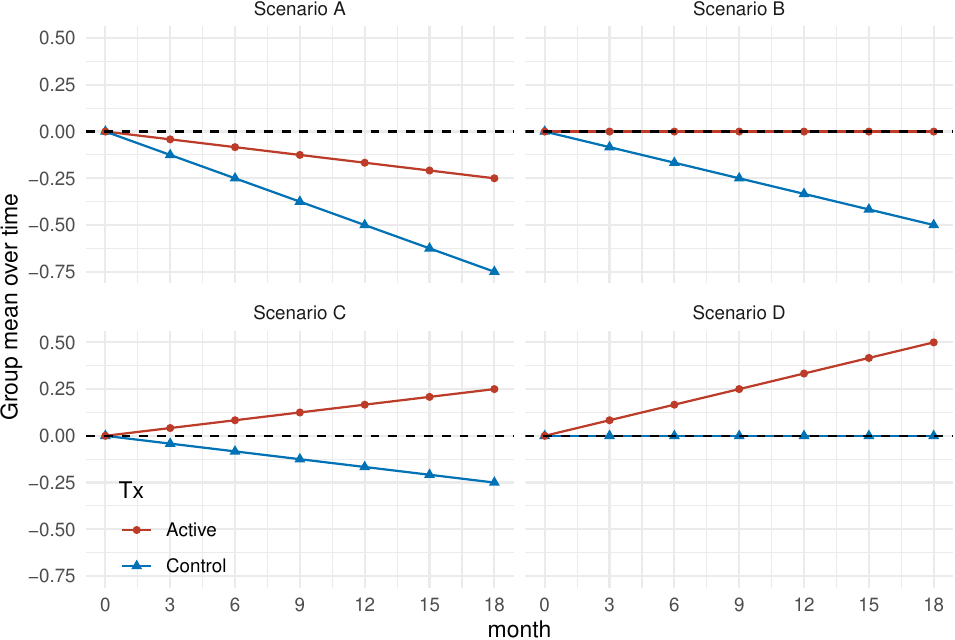}
\caption{\label{figlongmean} Four longitudinal mean trend scenarios assumed in simulations. All scenarios have the same linear effect (a difference between slopes of 0.5 units per 18 months). Scenarios A to C have finite proportional effects (2/3, 1, 2), while scenario D, with a control group mean fixed at zero, does not.}
\end{figure}

Figure \ref{figlongmean} illustrates the mean trends assumed for simulations. Each panel has the same linear treatment effect, a difference between slopes of 0.5 units per 18 months. Scenarios A to C have proportional effects of 2/3, 1, and 2. Scenario D does not have a finite proportional effect, because the control group mean is fixed at zero. Scenarios A to C satisfy the proportional effect assumption as well as a linear effect assumption. We simulate 10,000 trials under each of these scenarios with $N=200$ per group, residual variance 1.5, and random intercept variance 2. To each simulated dataset we fit two models, both of which assume an unstructured mean for the placebo group and either a linear treatment effect over time ("slope"), or a proportional treatment effect as defined in (\ref{eq1}). Both models correctly specify the temporal mean structure of the simulated data for scenarios A to C. Only the "slope" model is correctly specified for scenario D since it does not have a finite proportional effect. Both models include participant-specific random intercepts and constrain the group means at baseline to be equivalent (this is known as constrained Longitudinal Data Analysis (cLDA)). The models were fit by restricted maximum likelihood with the \texttt{lme} and \texttt{nlme} functions from R's \texttt{nlme} package  \citep{nlme}.

\begin{table}[t]
\begin{center}
\begin{minipage}{6in}
\caption{\label{tab1}Simulated power, Type I error, and proportion of rejections under the null that favor active treatment for the longitudinal mean trends shown in Figure \ref{figlongmean}. We conduct 10,000 simulations for each scenario, with $N=200$ per group, a residual variance of 1.5, and a random intercept variance of 2. For each simulated dataset, we fit two models: one assuming a linear treatment effect over time ("slope") and the other assuming a proportional treatment effect ("proportional") as defined in (\ref{eq1}). Both models assume an unstructured mean for the placebo group. Both models include participant-specific random intercepts and constrain the group means at baseline to be equivalent.} 
\begin{tabular}{lp{2.3cm}rrp{4.5cm}}
\toprule
Scenario & Treatment effect assumption & Power (\%) & Type I error (\%) & Proportion of rejections in favor of active under the null (\%) \\ 
\midrule
  A & Slope & 88.2 & 5.0 & 47.5 \\ 
     & Proportional & 93.8 & 5.3 & 93.2 \\ 
\midrule
  B & Slope & 88.7 & 5.0 & 51.3 \\ 
     & Proportional & 94.6 & 7.5 & 100.0 \\ 
\midrule
  C & Slope & 88.2 & 5.0 & 49.5 \\ 
     & Proportional & 78.1 & 17.7 & 100.0 \\ 
\midrule
  D & Slope & 88.6 & 5.2 & 50.2 \\ 
     & Proportional & 12.0 & 34.0 & 100.0 \\ 
\botrule
\end{tabular}
\end{minipage}
\end{center}
\end{table}

\begin{figure}
\centering
\includegraphics[width=6in]{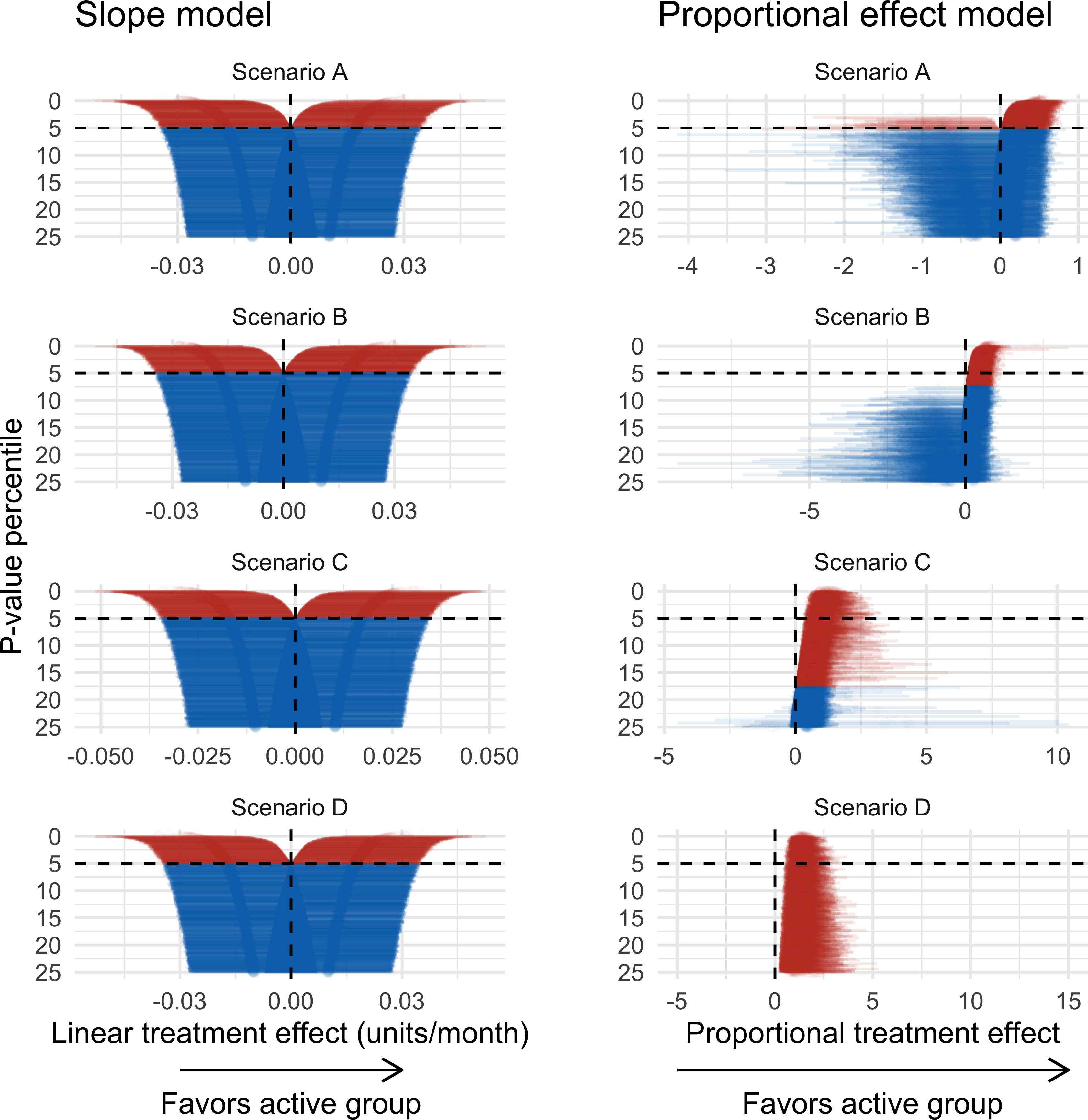}
\caption{\label{t1zipper}Zipper plots showing estimates and 95\% confidence intervals for longitudinal Type I error simulations. The intervals are sorted so that those associated with the smallest p-values are toward the top of each panel, and only the smallest 25\% of p-values are shown. Red intervals denote $p<0.05$. While the linear "slope" model estimates are symmetric about the zero (left), the plots for the proportional effect model (right) reveal bias in favor of treatment with a disproportionate number of positive dots. Confidence intervals also appear to be narrower when proportional effect estimates favor treatment. Vertical dashed lines are at the true value, zero. Horizontal dashed lines are at the target rejection rate of 5\%.}
\end{figure}

Table \ref{tab1} summarizes the longitudinal simulation results. The linear "slope" model has consistent operating characteristics for each scenario, as expected (88\% power, 5\% two-sided alpha, with about 50\% of rejections favoring active treatment under the null hypothesis). Meanwhile, the proportional effect model has decreasing power, and increasing Type I error as the control group mean approaches zero from A to D. The proportion of rejections favoring the active treatment under the null hypothesis is 93\% for scenario A and 100\% for the other scenarios. The zipper plots in Figure \ref{t1zipper} also demonstrate this bias in favor of treatment which is exacerbated as the control group mean approaches zero. Confidence intervals are also tighter when estimates favor active treatment.

\clearpage

\section{Discussion}\label{discussion}

These simulations demonstrate that the proportional effect parameterization promoted in the tutorial is biased when the control group mean is near zero, and identical to the linear model when the control group mean is sufficiently far from zero. This bias occurs even when the proportional effect assumption is met. The reported improved power of the longitudinal proportional effect model is likely erroneously inflated by bias in favor of active treatment. The proportional effect parameterization is sensitive to labeling of groups such that if we invert the group labels the model would be biased in favor of the control group. While the tutorial addresses testing the proportionality assumption, there is no mention of testing whether the control group is sufficiently far from zero in the tutorial or elsewhere.

As the power to detect treatment benefit is erroneously inflated, the power to detect treatment harm is erroneously deflated. Treatment emergent cognitive worsening is unfortunately a real possibility, as seen with the BACE inhibitor class of therapies for Alzheimer's disease. Worsening due to a BACE inhibitor was observed even with numeric improvement in the control group of a halted trial  \citep{sperling2021findings}. This harm is less likely to be detected by a proportional effect model.

Biostatisticians and clinical trialists should be aware of these fundamental issues with the tutorial's longitudinal proportional effect model and avoid it, particularly because of safety concerns. Bias corrections like those for ratio estimation with paired data  \citep{tin1965comparison} are conceivable. But it stands to reason that if the model's bias could be corrected, power would align with more flexible linear model alternatives and lose some, if not all, of its appeal. Uncertainty intervals for proportional effects and time savings  \citep{raket2022progression} can always be examined based on flexible linear models with the delta method  \citep{wang2024novel}, or the bootstrap, or with Markov chain Monte Carlo credible intervals.


\section{Supplemental material}

R code to reproduce simulations is available from \url{https://github.com/mcdonohue/PropEffects}

\section{Funding}

This work is supported in part by funding from the National Institute on Aging of the National Institutes of Health (U24AG057437).

\section{Conflict of interest}

None declared.

\bibliographystyle{abbrvnat}
\bibliography{reference}

\end{document}